\documentclass[sigconf]{acmart}

\usepackage{booktabs} 
\usepackage{balance}
\usepackage{todonotes}
\usepackage{subcaption}
\usepackage{listings}
\usepackage{color}

\definecolor{BrickRed}{RGB}{182, 50, 28}
\definecolor{commentcolor}{rgb}{0,0.6,0}
\definecolor{linenumbercolor}{rgb}{0.5,0.5,0.5}
\definecolor{stringliteralcolor}{rgb}{0.58,0,0.82}
\lstnewenvironment{code}[1][]%
  {\noindent\minipage{\linewidth}\medskip 
   \lstset{basicstyle=\ttfamily\footnotesize,frame=single,#1,escapeinside={(*@}{@*)}}}
  {\endminipage}

\setcopyright{rightsretained}

\acmDOI{10.475/123_4}

\acmISBN{123-4567-24-567/08/06}

\acmConference[CIKM'18]{CIKM 2018 International Conference on Information and Knowledge Management}{October 2018}{Lingotto, Turin, Italy}
\acmYear{2018}
\copyrightyear{2018}

\begin{document}
\title{MonetDBLite: An Embedded Analytical Database}

\author{Mark Raasveldt}
\affiliation{%
  \institution{CWI}
  \streetaddress{Science Park 123}
  \city{Amsterdam, Netherlands}
}
\email{m.raasveldt@cwi.nl}

\author{Hannes M\"uhleisen}
\affiliation{%
  \institution{CWI}
  \streetaddress{Science Park 123}
  \city{Amsterdam, Netherlands}
}
\email{hannes@cwi.nl}


\begin{abstract}
While traditional RDBMSes offer a lot of advantages, they require significant effort to setup and to use. Because of these challenges, many data scientists and analysts have switched to using alternative data management solutions. These alternatives, however, lack features that are standard for RDBMSes, e.g. out-of-core query execution. In this paper, we introduce the embedded analytical database MonetDBLite. MonetDBLite is designed to be both highly efficient and easy to use in conjunction with standard analytical tools. It can be installed using standard package managers, and requires no configuration or server management. It is designed for OLAP scenarios, and offers near-instantaneous data transfer between the database and analytical tools, all the while maintaining the transactional guarantees and ACID properties of a standard relational system. These properties make MonetDBLite highly suitable as a storage engine for data used in analytics, machine learning and classification tasks.
\end{abstract}

%
%



\maketitle

\section{Introduction}\label{section:introduction}

Modern machine learning libraries and analytical tools have moved away from using traditional relational databases for managing data. Instead, data is managed by storing it either as structured text (such as CSV or JSON files), or as binary files such as Parquet or HDF5~\cite{HDF5, PARQUET}. Managing data in flat files requires significant manual effort to maintain and is difficult to reason about because of the lack of a rigid schema. Furthermore, data managed in such a way is prone to data corruption because of lack of transactional guarantees and atomic write actions.

Relational database systems have been designed specifically to solve these problems. However, data scientists still prefer to use flat files. This is primarily because the current methods of using a relational database in combination with these analytical tools are lacking. The standard approach of using a database with an analytical tool is to run the database system as a separate process (the ``database server'') and connecting with it over a socket using a client interface (typically ODBC or JDBC). However, not only is this approach very slow~\cite{protocolpaper}, it is very cumbersome as the database system must be installed, managed and tuned as a separate process. For many use cases, the added effort of using a relational database outweighs the benefits of using one. 

Instead of managing the database as a separate process, the database can run embedded inside the analytical tool. This method has the advantage that the database server no longer needs to be managed, and the database can be installed from within the standard package manager of the tool. In addition, because the database and the analytical tool run inside the same process on the same machine, data can be transferred between them for a much lower cost.

SQLite~\cite{sqlite} is the most popular embedded database. It has bindings for all major languages, and it can be embedded without any licensing issues because its source code is in the public domain. However, it is first and foremost designed for transactional workloads on small datasets. While it can be used in conjunction with popular analytical tools, it does not perform well when used for analytical purposes.


In this paper, we describe MonetDBLite, an Open-Source embedded database based on the popular columnar database MonetDB~\cite{monetdb}. MonetDBLite is an in-process analytical database that can be run directly from within popular analytical tools without any external dependencies. It can be installed through the default package managers of popular analytical tools, and has bindings for C/C++, R, Python and Java. Because of its in-process nature, data can be transferred between the database and these analytical tools at zero cost. The source code for MonetDBLite is freely available\footnote{\url{https://github.com/hannesmuehleisen/MonetDBLite}} and is in active use by thousands of analysts around the world.

\textbf{Contributions.} The main contributions of this paper are:

\begin{itemize}
\item We describe the internal design of MonetDBLite, and how it interfaces with standard analytical tools.
\item We discuss the technical challenges we have faced in converting a popular Open-Source database into an in-process embeddable database.
\item We benchmark MonetDBLite against other alternative database systems when used in conjunction with analytical tools, and show that it outperforms alternatives significantly. This benchmark is completely reproducible with publicly available sources.
\end{itemize}

\begin{figure*}[!ht]
    \centering
    \begin{subfigure}[t]{.32\textwidth}
        \centering
        \includegraphics[height=.6\textwidth]{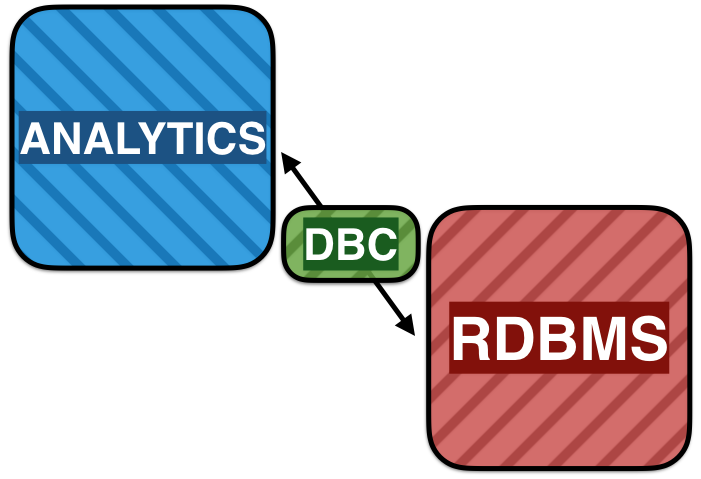}
        \caption{Socket connection.}
    \end{subfigure}
    \begin{subfigure}[t]{.32\textwidth}
        \centering
        \includegraphics[height=.6\textwidth]{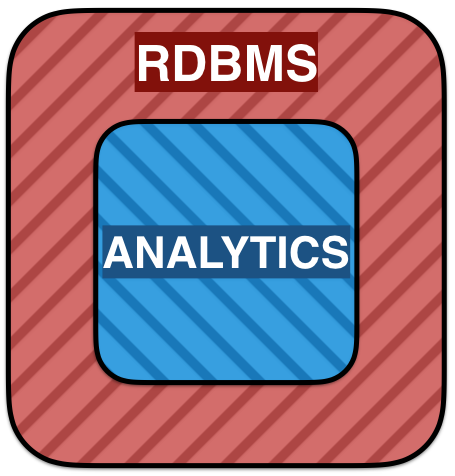}
        \caption{In-database processing.}
    \end{subfigure}
    \begin{subfigure}[t]{.32\textwidth}
        \centering
        \includegraphics[height=.6\textwidth]{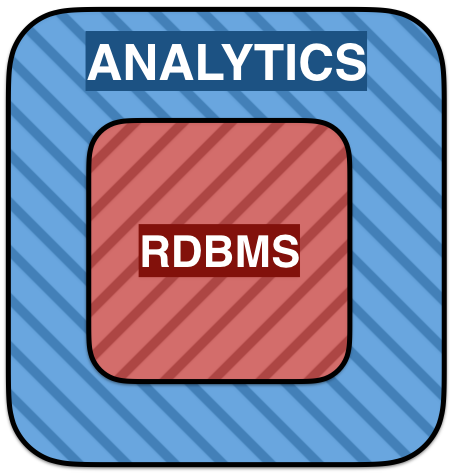}
        \caption{Embedded database.}
    \end{subfigure}
    \caption{Different ways of connecting analytical tools with a database management system.}
    \label{image:databaseanalytical}
\end{figure*}

\textbf{Outline.} This paper is organized as follows. In Section~\ref{section:relatedwork}, we discuss related work. In Section~\ref{section:designandimplementation}, we describe the design and implementation of the MonetDBLite system. We compare the performance of MonetDBLite against other database systems and statistical libraries in Section~\ref{section:evaluation}. Finally, we draw our conclusions in Section~\ref{section:conclusion}.

\section{Related Work}\label{section:relatedwork}

There are three main methods of combining relational databases with analytical tools. These methods are visualized in Figure~\ref{image:databaseanalytical}. The standard approach is to connect the analytical tool to a database through a database connection (DBC). The client can then issue queries to the database over this client connection and retrieve any data stored in the database. This approach is database agnostic, and allows the user to stay within their familiar scripting language environment (REPL or IDE). However, exporting large amounts of data from the database to the client is very inefficient~\cite{protocolpaper}, and can be a significant bottleneck in analysis pipelines.

An alternative solution is to use in-database processing methods. By executing the analysis pipelines inside the database, the overhead of data export can be entirely avoided. In addition, in-database processing can provide additional benefits in the form of automatic parallelization or distributed execution of these functions~\cite{pythonudfs} within the database engine.

While this approach removes the data transfer overhead between the scripting language and the database, it still requires the user to run and manage a separate database server. These user-defined functions also introduce new issues. Firstly, they force users to rewrite code so the code fits within the query work flow. Secondly, because these UDFs run within the database process, they are difficult to create and debug~\cite{debuggingudfs}. Users cannot use the IDEs/REPLs/debugging tools that they are familiar with (such as RStudio or PyCharm) while writing the user-defined functions. Additionally, user-defined functions introduce safety issues as arbitrary code can now run within the database kernel.

Instead of running the database system as a separate server, it can be embedded into the scripting language. The database can be accessed using the same database-agnostic interface as a standard database client connection, but because the database resides in the same address space as the scripting language, data can be transferred between the two systems with significantly less overhead.

Another advantage of using an embedded database inside analytical scripts is that the embedded database does not introduce any environment dependencies, making the analytical script portable and capable of running anywhere without additional user effort. Using a standard relational database requires the user to have the database server running in the background or on a separate machine, whereas the embedded database can be completely controlled from within the script itself. 

Embedded databases are extremely popular, mainly because of the omnipresent SQLite~\cite{sqlite}. This embedded database is shipped with all major operating systems and browsers, and is included with numerous other applications. SQLite has become so ubiquitous because it provides the valuable advantages of a relational database management system for a low development cost. It has bindings for all major languages, and it can be embedded without any licensing issues because its source code is in the public domain. According to the author of SQLite, there are over one trillion SQLite databases in active use.

However, SQLite is first and foremost designed for transactional workloads. It is a row-store database that uses a volcano-style processing model for query execution. While popular analytical tools such as Python and R do have SQLite bindings, it does not perform well when used for analytical purposes. Even exclusively using SQLite as a storage engine typically does not work out well in these scenarios as it is not optimized for bulk retrieval of data. In addition, often only a limited subset of columns of a table are used in analyses, and its row-wise storage layout forces it to always scan entire tables. This can leads to poor performance when dealing with wide data.

There are also alternatives to using a relational database system at all. Libraries such as dplyr~\cite{dplyr}, data.table~\cite{datatable} and Pandas~\cite{pandas} allow the user to execute common database operations directly from within analytical scripting languages. They implement operations such as joins and aggregations that efficiently operate directly on the native structures used in these languages. These solve the problem of users wanting to execute database operations, however, they do not solve the problem of automatically managing persistent data for the user. In addition, they do not facilitate efficient out-of-memory execution.


\section{Design \& Implementation}\label{section:designandimplementation}
In this section we will discuss the general design and implementation of MonetDBLite, and the design choices we have made while implementing it.

\subsection{Internal Design}\label{subsection:internaldesign}
MonetDBLite is based on the popular Open-Source columnar database MonetDB, and as such it shares most of its internal design. The core design of MonetDB is described in Idreos et al.~\cite{monetdb}. However, since this publication a number of core features have been added to MonetDB. In this section, we give a brief summary of the internal design of MonetDB and describe the features that have been added to MonetDB since.

\textbf{Data Storage.} MonetDBLite stores relational tables in a columnar fashion. Every column is stored either in-memory or on-disk as a tightly packed array. Row-numbers for each value are never explicitly stored. Instead, they are implicitly derived from their position in the tightly packed array. Missing values are stored as ''special`` values within the domain of the type, i.e. a missing value in an \texttt{INTEGER} column is stored internally as the value $-2^{31}$.

Columns that store variable-length fields, such as CLOBs or BLOBs, are stored using a variable-sized heap. The actual values are inserted into the heap. The main column is a tightly packed array of offsets into that heap. These heaps also perform duplicate elimination if the amount of distinct values is below a threshold; if two fields share the same value it will only appear once in the heap. The offset array will then point to the same heap entry for the rows that share the same value.

\textbf{Memory Management.} MonetDB does not use a traditional buffer pool to manage which data is kept in memory and which data is kept on disk. Instead, it relies on the operating system to take care of this by using memory-mapped files to store columns persistently on disk. The operating system then loads pages into memory as they are used and evicts pages from memory when they are no longer being actively used. This model allows it to keep hot columns loaded in memory, while columns that are not frequently touched are off-loaded to disk. 


\textbf{Concurrency Control.} MonetDB uses an optimistic concurrency control model. Individual transactions operate on a snapshot of the database. When attempting to commit a transaction, it will either commit successfully or abort when potential write conflicts are detected.

\textbf{Query Plan Execution.} SQL is first parsed into a relational algebra tree and then translated into an intermediate language called MAL (Monet Assembly Language). MAL instructions process the data in a column-at-a-time model. Each MAL operator processes the full column before moving on to the next operator. The intermediate values generated by the operators are kept around in-memory if not too large, and passed on to the next operator in the pipeline.

Optimizations happen at three levels. High level optimizations, such as filter push down, are performed on the relational tree. Afterwards, the MAL code is generated and further optimizations are performed such as common sub-expression elimination. Finally, during execution tactical decisions are made about how specific operations should be executed, such as which join implementation to use.

\begin{figure}[t]
\begin{center}
    \includegraphics[width=\columnwidth]{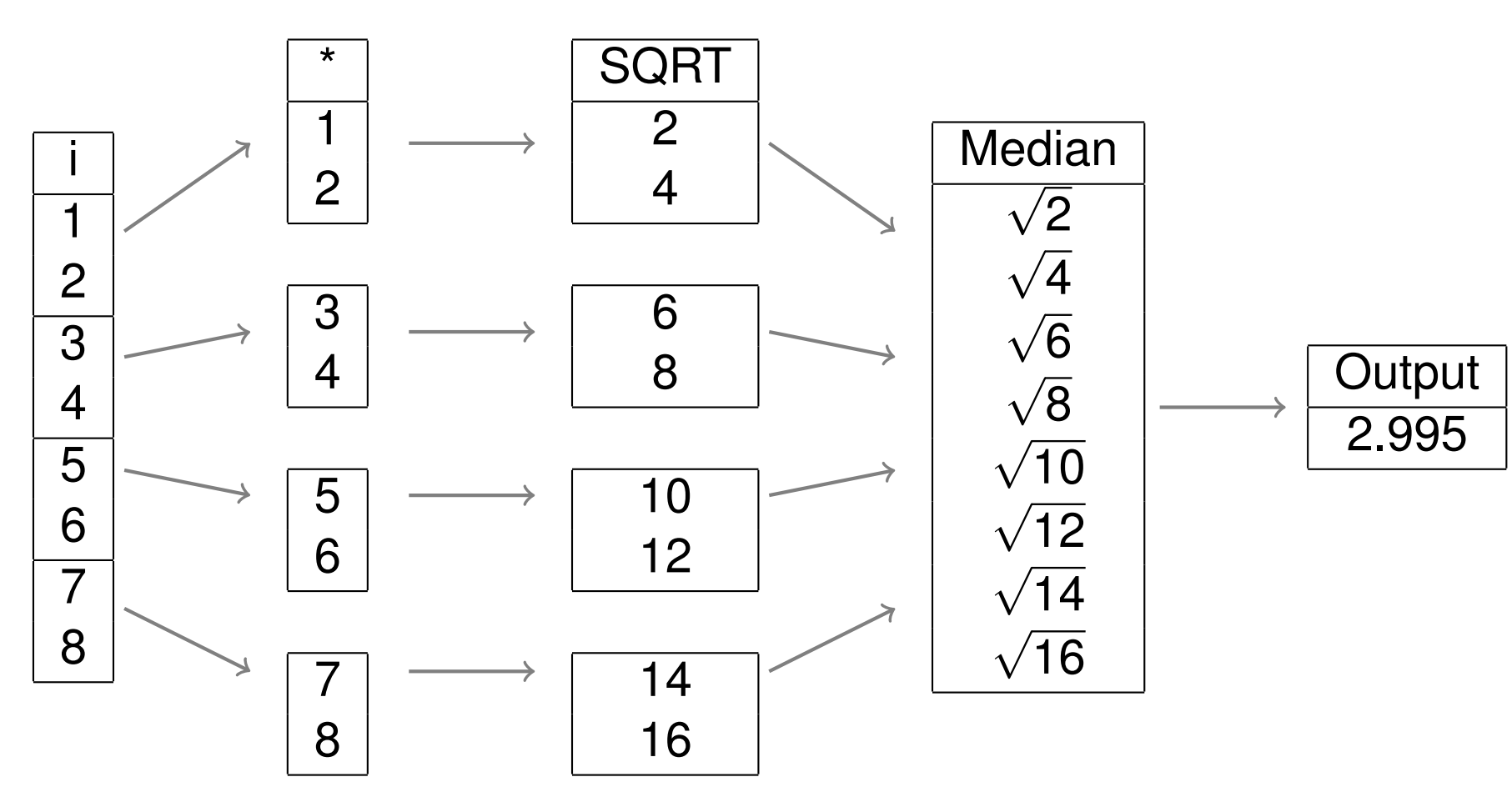}
    \caption{Parallel execution in MonetDB.}
    \label{image:parallelchain}
\end{center}
\end{figure}

\textbf{Parallel Execution.} Initially, a sequential execution plan is generated. Parallelization is then added in the second optimization phase. The individual MAL operators are marked as either ``blocking'' or ``parallelizable''. The optimizers will alter the plan by splitting up the columns of the largest table into separate chunks, then executing the ``parallelizable'' operators once on each of the chunks, and finally merging the results of these operators together into a single column before executing the ``blocking'' operators. This is visualized in Figure~\ref{image:parallelchain} for the query \texttt{SELECT MEDIAN(SQRT(i * 2)) FROM tbl}.

The amount of chunks that are generated is decided by a set of heuristics based on base table size, the amount of cores and the amount of available memory. The database will attempt to generate chunks that fit inside main memory to avoid swapping, and will attempt to maximize CPU utilization. In addition, the optimizer will not split up small columns as the added overhead of parallel execution will not pay off in this case.

\textbf{Automatic Indexing.} In addition to allowing the user to manually build indices through the \texttt{CREATE INDEX} commands, MonetDB will automatically create indices during query execution.

Imprints~\cite{imprints} are a bitmap index that are used to assist in efficiently computing point and range queries. The bitmap index holds, for each cache line, a bitmap that contains information about the range of values in that cache line. They are automatically generated for persistent columns when a range query is issued on a specific column. They are then persisted on disk and used for subsequent queries on that column. Imprints are destroyed when a column is modified.

Hash tables are also automatically created for persistent columns when they are used in groupings or as join keys in equi-joins. These are also persisted on disk. Hash tables are destroyed on updates or deletions to the column. Unlike imprints, however, they are updated on appends to the tables.

\textbf{Order Index.} In addition to imprints and hash tables, MonetDB supports creation of a sorted index that is not created automatically. It must be created using the \texttt{CREATE ORDER INDEX} statement. Internally, the order index is an array of row numbers in the sort order specified by the user. The order index is used to speed up point and range queries, as well as equi-joins and range-joins. Point and range queries are answered by using a binary search on the order index. For joins, the order index is used for a merge join.

\subsection{Embedding Interface}
MonetDBLite is a database that is embedded into analytical tools directly, rather than running as a standard client-server database. As MonetDBLite runs within a process, clients have to create and initialize the database themselves rather than connecting to an existing database server through a socket connection. For this purpose, MonetDBLite needs a set of language bindings so the database can be initialized and queries can be issued to the database.

MonetDBLite has language bindings for the C/C++, R, Python and Java programming languages. However, all of these are wrappers for the C/C++ language bindings. The main challenge in creating these wrappers is converting the data to and from the native types of each of these languages. The optimization challenges of this type conversion are discussed in Section~\ref{subsection:nativelanguage}. In this section, we will discuss only the C/C++ API. 

The database can be initialized using the \texttt{monetdb\_startup} function. This function takes as optional parameter either a reference to a directory in which it can persistently store any data. If no directory is provided, MonetDBLite will be launched in an in-memory only mode, in which case no persistent data is saved to disk.

If the database is launched in persistent mode, a new database will be created in the specified directory if none exists yet. Otherwise, the existing database will be loaded and potentially upgraded if it was created by an older version of MonetDBLite. If the database is launched in-memory, a new temporary database will be created that will be kept entirely in-memory. Any data added to the database will be kept in-memory as well. After an in-memory database is shut down, all stored data will be discarded. The regular MonetDB does not have this feature.

After a database has been started, connections to the database can be created using the \texttt{monetdb\_connect} function. In the regular MonetDB server, these connections represent socket connections to a client process. In MonetDBLite, however, these connections are dummy clients that only hold a query context and can be used to query the database. Multiple connections can be created for a single database instance. These connections can be used for inter-query parallelism by issuing multiple queries to the database in parallel and they provide transaction isolation between them.

Using these connections, the embedded process can issue standard SQL queries to the database using the \texttt{monetdb\_query} function. This function takes as input a client context and a query to be issued, and returns the results of the query to the client in a columnar format in a \texttt{monetdb\_result} object. The \texttt{monetdb\_result} object is semi-opague, exposing only a limited amount of header information, as shown in Listing~\ref{lst:monetdbresult}

\begin{code}[frame=none,language=C,caption={MonetDBLite Result Object},label=lst:monetdbresult]
struct monetdb_result {
	size_t nrows;
	size_t ncols;
	char type;
	size_t id;
};
\end{code}

The individual columns of the result can be fetched using the \texttt{monetdb\_result\_fetch} function, which takes as input a pointer to the \texttt{monetdb\_result} object and a column number. There are two versions of this function: a low level version, and a high level version. In the low level version, the underlying structures used by the database are directly returned without any conversions being performed. This function requires internal knowledge of the database internals, and is intended for use primarily for the language-specific wrappers for extra performance. In the high level version, the database structures are converted into a set of simple structures that can be used without knowledge of the internals of MonetDB(Lite). The returned structures depend on the type of the column. An example for the \texttt{int} type is given in Listing~\ref{lst:monetdbcolumn}.

\begin{code}[frame=none,language=C,caption={MonetDBLite Integer Column},label=lst:monetdbcolumn]
struct monetdb_column {
		monetdb_type type;
		int* data;
		size_t count;
		int null_value;
		double scale;
		int (*is_null)(int value);
};
\end{code}

In addition to issuing SQL queries, the embedded process can efficiently bulk append large amounts of data to the database using the \texttt{monetdb\_append} function. This function takes the schema and the name of a table to append to, and a reference to the data to append to the columns of the table. This function allows for efficient bulk insertions, as there is significant overhead involved in parsing individual \texttt{INSERT INTO} statements, which becomes a bottleneck when the user wants to insert a large amount of data.

\subsection{Native Language Interface}\label{subsection:nativelanguage}
For any of the languages other than C/C++, data has to be converted between the database's native format to the target language's native format. When a SQL query is issued, the result has to be mapped back into the target environment. Likewise, if the user wants to move data from the target environment to the database, it has to be converted. 

Database connectors in the target environment face a similar but more difficult problem, as they also have to deal with communicating with the remote database server. We could adapt these database connectors to work with MonetDBLite. However, in an analytical context this approach is problematic. As these are row-focused interfaces~\cite{protocolpaper}, the results of queries must be fetched one-by-one. This leads to a large amount of overhead when fetching a large result set, especially in interpreted scripting languages such as R or Python. Columnar bulk access to result sets is therefore needed, where all values belonging to one column can be fetched into a set of arrays, one per column, in one or few calls to the database interface. 




However, not all arrays are created equal. While it is possible to subclass the native array representation in most programming environments, efficiency concerns and expectations by third-party software might make a fully native data representation necessary. For example, in R, most third-party packages will contain some portion of compiled code written in C/C++, which relies on arrays being stored in the native bit representation if they are to compute anything meaningful with them. Similarly, in the NumPy environment, third-party packages can get a pointer to the native C representation of any array. Hence for the objects that we return from the database to be able to be used by these packages, we must create objects that exactly match the native array format of the target environment. 

\textbf{Zero-Copy.}  Every target environment has a particular array representation in memory. However, due to hardware support contiguous C-style arrays are ubiquitous for numerical values. For example, both R and NumPy use this representation to store arrays of numerical data. This allows for a unique optimization opportunity: Instead of converting the data into a freshly allocated memory area, we can choose to share a pointer to the existing data with the target system. The memory layout needs to be compatible between data management and target environment, e.g. both using contiguous C-style arrays containing four-byte signed integers. If this pointer sharing is possible, the only cost comes from initializing metadata structures in the target environment (e.g. SEXP header in R). However, this cost does not depend on the size of the data set. 

Great care needs to be taken to prevent modification of the data being shared. The target environment may run any program imaginable, including code from contributed packages, that may try to modify the shared memory areas. The shared pointer might be part of persistent data of the database, hence modifying the data directly could lead to corruption of the data stored in the database. Because of this, no direct modification of this data is allowed. 

What is desirable here are copy-on-write semantics for the shared data. If code from the target environment attempts to write into the shared data area, the data should be copied within the target environment and only the copy modified. To ensure these semantics are enforced, the Unix \texttt{mprotect} kernel function can be used to disallow writes to the data by the target code. When the target environment attempts to modify data we have shared with it, we create copy and modify the copy instead. This allows for efficient read-only access without the risk of data corruption.

\begin{figure}[!h]
    \includegraphics[width=.8\columnwidth]{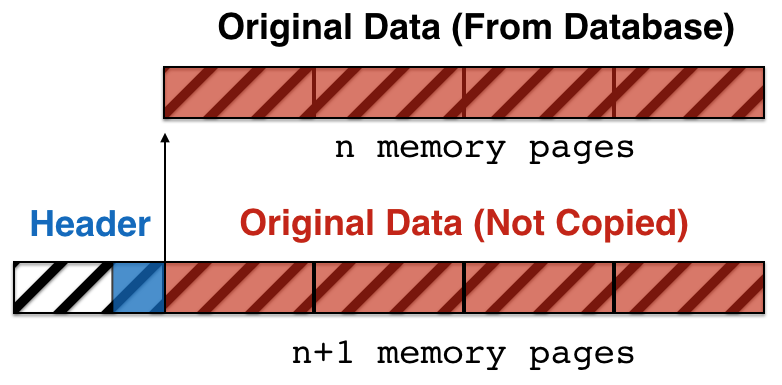}
    \caption{Header forgery for zero-copy data transfer.}
    \label{fig:headerforgery}
\end{figure}

\textbf{Header Forgery.} A challenge of providing a zero-copy interface to the data stored in the database is that certain libraries expect metadata to be stored as a header physically in front of the data. This is accomplished in the library by performing a single memory allocation that allocates the size of the header plus the size of the data. This is problematic in our scenario. As the source data comes directly from an external database system it does not have space allocated in front of it for these headers. 

This problem could be solved by making the database always allocate extra bytes in front of any data that could be passed to the analytical tool. As we have full control of the database system, this is feasible. However, it would require a significant amount of code modification and would result in wasted space in scenarios where the data is not passed to the analytical tool.

Instead, we solve this problem using header forgery. This process is shown in Figure~\ref{fig:headerforgery}. To provide a zero-copy interface of a region of \texttt{n} memory pages, we allocate a region of size $n+1$ memory pages using the \texttt{mmap}~\cite{mmap} function. We then place the header information at the end of the first page. We then use the \texttt{mmap} function together with the \texttt{MAP\_FIXED} flag to directly link the remaining \texttt{n} memory pages to the original data. This linking happens in the memory page table, and does not create a copy of these pages. This method predates, but could be considered an application of the memory rewiring technique presented in~\cite{RUMA}.

\begin{figure}[!h]
    \includegraphics[width=.8\columnwidth]{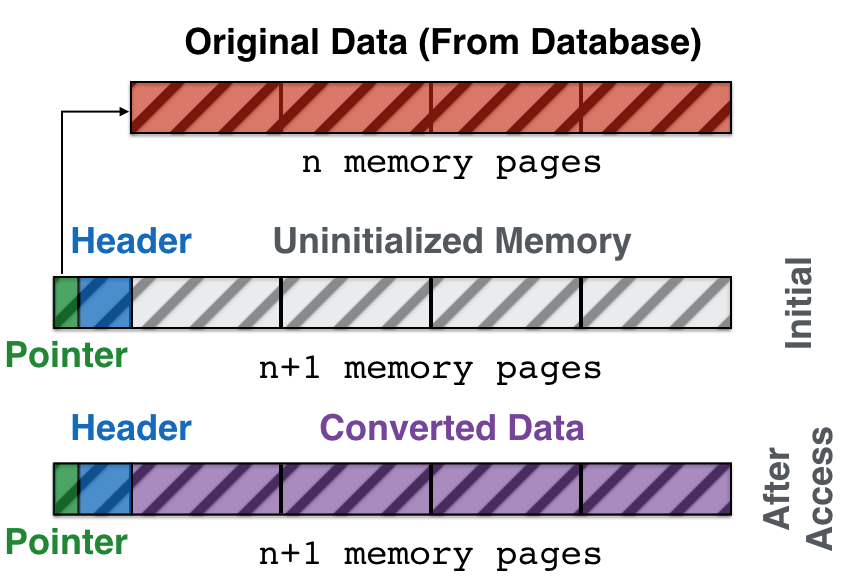}
    \caption{Lazy data conversion.}
    \label{fig:lazyconversion}
\end{figure}

\textbf{Lazy Conversion.} While the zero-copy approach is ideal, as it does not require us to touch the to--be--converted data, it cannot be used in all cases. When the internal representation of the database is not bit-compatible with that of the target environment, data conversion has to be performed. As all data has to be converted, the conversion will take a linear amount of time w.r.t. the size of the result set. However, it is not known whether the target environment will ever actually do anything with the converted data. It is not uncommon for a user to perform a query such as \texttt{SELECT * FROM table} and only access a small amount of columns from the result.

This issue can be resolved by performing lazy conversion of the result set. Instead of eagerly converting the entire result set, we create a set of ``dummy'' arrays that start out with a correctly initialized header. However, the data is filled with uninitialized memory. This is shown in Figure~\ref{fig:lazyconversion}. We then use the \texttt{mprotect}~\cite{mprotect} function to protect the uninitialized memory from being read or written to directly using the \texttt{PROT\_NONE} flag. When the user attempts to access the protected memory area, the system throws a segmentation fault, which we then catch using a signal handler. Using a pointer to the original data that is stored alongside the header, we then perform a conversion of the actual data and unset the \texttt{mprotect} flag, allowing the user to use the now-converted data transparently.

\subsection{Technical Challenges}\label{section:technicalchallenges}
In this section, we will discuss the additional technical challenges that we encountered while converting a standard relational database management system to an in-process embedded database system.

\textbf{Internal Global State.} MonetDB was originally designed to run as a single stand-alone process. One of the consequences of this design is that internal global state (global variables) is used often in the source code. The database uses global state to keep track of e.g. the data stored inside the database, the write-ahead logger and numerous database settings. 

This global state leads to a limitation: it is not possible to run MonetDBLite twice in the same process. As the global state holds all the information necessary for the database to function, including paths to database files, and this information is continuously accessed while the database is running, only one database server can be running in the same process. To make it possible to run several database servers within the same process would require a very comprehensive code rewrite, as the global database state would have to be passed around to almost every function.

\textbf{Garbage Collection.} Another issue caused by this global state is garbage collection. As the database server no longer runs as a stand-alone program, the global variables can no longer be reset by restarting the server. In addition, all the allocated memory has to be freed in the process. Allocated regions can no longer be neglected with the knowledge that they will be freed when the process is terminated. Instead, to properly support an ``in-process shutdown'' of the database server, everything has to be cleaned up manually and all global variables have to be reset to their initial state. 

\textbf{External Global State.} Another consequence of the database server being designed to run as a stand-alone process is that it modifies a lot of external global state, such as signal handlers, locale settings and input/output streams. For each of these, it was necessary to modify the database source code to not modify the global state. Otherwise loading the database package would result in it overriding signal handlers, leading to e.g. breaking the scripting languages' input console.

Calls to the \texttt{exit} function were especially problematic. In the stand-alone version of MonetDB the database server shuts down when a fatal error was detected (such as running with insufficient permissions or attempting to open a corrupt database). This happens mostly during start-up. This is expected behavior in a stand-alone database server, but becomes problematic when running embedded inside a different program. Attempting to access a corrupt database using the embedded database would result in the entire program crashing, rather than a simple error being thrown. Even worse, since the database would simply exit in these scenarios, no alternative path exists to only report the error. To avoid a large code rewrite, we used \texttt{longjmp} whenever the \texttt{exit} function was called, which would jump out of the \texttt{exit} and move to a piece of code where the error could be reported. 

\textbf{Error Handling.} Another aspect of the database design that we needed to rethink was error handling. In the regular database server, errors are reported by writing them to the output stream so they can be handled by the client program. However, in the embedded version the errors must be reported as a return value from the SQL query function. We had to rewrite large portions of the error reporting code to accommodate this.


\textbf{Dependencies.} To make MonetDBLite as simple to install as possible, one of our design goals was to remove all external dependencies. Regular MonetDB has a large number of required dependencies, among which are \texttt{pcre}, \texttt{openssl}, \texttt{libxml} and \texttt{pkg-config} along with a large number of optional dependencies. For MonetDBLite, we stripped all of these dependencies by removing large chunks of optional code and rewriting code that relied on any of the required dependencies. For example, we made our own implementation of the \texttt{LIKE} operator (that previously used regular expressions from the PCRE library). As a result of our efforts, MonetDBLite has no external dependencies and can be installed without having to install any other libraries. 


















\section{Evaluation}\label{section:evaluation}
In this section, we perform an evaluation of the performance of MonetDBLite and compare it against both (1) other relational database management systems, and (2) several popular RDBMS alternatives used in statistical tools.

\subsection{Setup}
All experiments in this section were run on a desktop-class computer with an Intel i7-2600K CPU clocked at 3.40GHz and 16 GB of main memory running Fedora 26 Linux with Kernel version 4.14. We used GCC version 7.3.1 to compile systems. Reported timings are the median of ten hot runs. The initial cold run is always ignored. A timeout of 5 minutes is used for the queries.

\textbf{Systems.} The following systems were used to compare against in our benchmarks. All systems were configured to only use one of the eight available hardware threads for fairness (as not all systems support intraquery parallelism). Furthermore, unless indicated otherwise, we have attempted to configure the systems to take full advantage of available memory. The complete configuration settings and scripts to reproduce the results reported below can be found in the benchmark repository\footnote{\url{https://github.com/Mytherin/MonetDBLiteBenchmarks}}.

\begin{itemize}
\item \textbf{SQLite}~\cite{sqlite} (Version 3.20.1) is an embedded SQL database designed for transactional workloads.
\item \textbf{MonetDB}~\cite{monetdb} (Version 11.29.3) is an analytical column-store database.
\item \textbf{PostgreSQL}~\cite{postgresql} (Version 9.6.1) is a row-store database designed for transactional workloads.
\item \textbf{MariaDB}~\cite{MYSQL} (Version 10.2.14) is a row-store database designed for transactional workloads. It is based on the popular MySQL database.
\end{itemize}

\textbf{Libraries.} In addition to the above-mentioned database management systems, we test the following analytical libraries that emulate database functionality. We only use these libraries in the query execution benchmarks.

\begin{itemize}
\item \textbf{data.table}~\cite{datatable} (Version 1.11.0) is an R library for performing common database operations.
\item \textbf{dplyr}~\cite{dplyr} (Version 0.7.4) is an R library for performing common database operations.
\item \textbf{Pandas}~\cite{pandas} (Version 0.22.0) is a Python library for performing common database operations.
\item \textbf{Julia}~\cite{julia} (Version 0.6.2) is a JIT compiled analytical language that has support for performing standard database operators through the \texttt{DataFrames.jl} library.
\end{itemize}

\textbf{Datasets.} We perform benchmarks using the following data sets.

\begin{itemize}
\item \textbf{TPC-H Benchmark.}~\cite{tpch}. This synthetic dataset is designed to be similar to real-world data warehouse fact tables. In our benchmarks, we use the scale factors 1 and 10. The scale factor indicates approximately the size of the dataset in GB.

\item \textbf{American Community Survey (ACS)}~\cite{ACSDATA}. This dataset contains millions of census survey responses. It consists of 274 columns. 
\end{itemize}

\subsection{TPC-H Benchmark}
As we focus on the integration of analytical tools with an analytical database, there are three different scenarios that we want to optimize for and that we will benchmark. 

\begin{enumerate}
    \item \textbf{Data Ingestion.} The rate at which data can be imported into the database from the analytical tool. We call this the data ingestion or data import rate. This scenario occurs when users want to take data that is the result of computations in the analytical tool and store it persistently in the database.
    \item \textbf{Data Export.} The rate at which data can be imported into the analytical tool from the database. This data export rate is important when the user wants to perform analytics on data that is stored persistently within the RDBMS. 
    \item \textbf{Query Execution.} The performance of the database engine when performing analytical queries. This scenario occurs when the user wants to perform operations and aggregations on large amounts of data using the databases' storage engine. Note that for query execution, it is also possible to simply move the data from the database into the analytical tool and do the processing there using the previously mentioned libraries. For that reason, we also compare the performance of the RDBMS with the afore-mentioned libraries. 
\end{enumerate}

\subsection*{Data Ingestion}
\begin{figure}[!t]
    \includegraphics[width=.8\columnwidth]{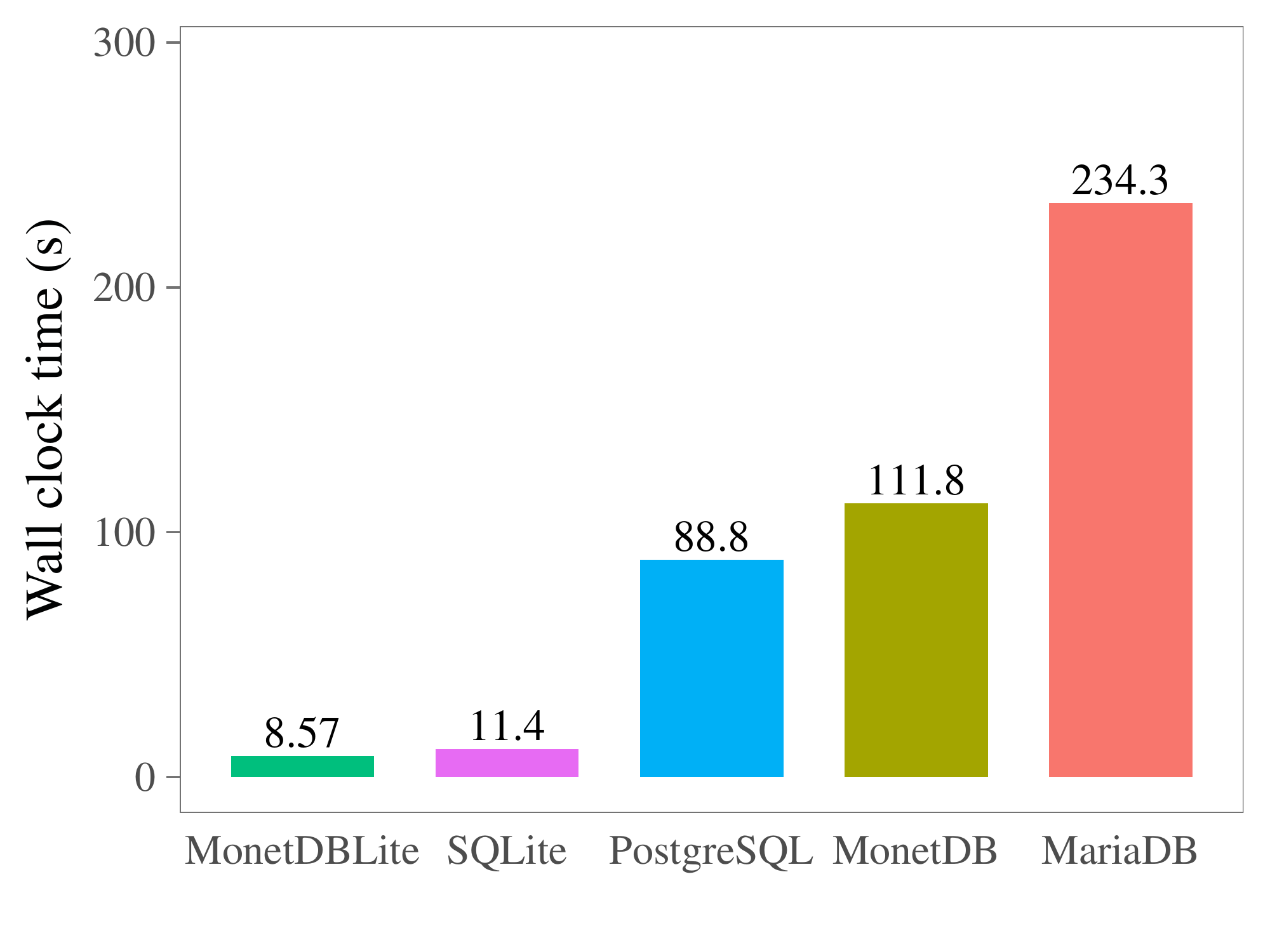}
    \caption{Writing the \texttt{lineitem} table from R to the database.}
    \label{fig:dataingestion}
\end{figure}

For the data ingestion benchmark, we only consider the \texttt{lineitem} table. This is the biggest table in TPC-H. It has 16 columns, primarily of types \texttt{DECIMAL}, \texttt{DATE} and \texttt{VARCHAR}. There are no \texttt{NULL} values.

For this experiment, we read the entire \texttt{lineitem} table into R and then use the \texttt{dbWriteTable} function of the R DBI~\cite{dbi} API to write the table into the database. After this function has been completed, the table will be persistently present within the database storage engine and all the data will have been loaded into the database. We only consider the database systems for this experiment.

The results of this experiment can be seen in Figure~\ref{fig:dataingestion}. We can see that MonetDBLite has the fastest data ingestion. However, we note that SQLite is not very far behind MonetDBLite. For both systems, the primarily bottleneck is writing the data to disk. MonetDBLite gains performance by storing the data in a more compact columnar format, rather than the B-tree structure that SQLite uses to store data internally.

All the other systems perform extremely poorly on this benchmark. This is because the data is written to the database over a socket connection, which requires a large amount of network communication. However, the main problem is that these database systems do not have specialized protocol code for copying large amounts of data from the client to the server console. Instead, the data is inserted into the database using a series of \texttt{INSERT INTO} statements, which introduces a large amount of overhead leading to orders of magnitude worse performance than the embedded database systems.

\subsection*{Data Export}
\begin{figure}[!t]
    \includegraphics[width=.8\columnwidth]{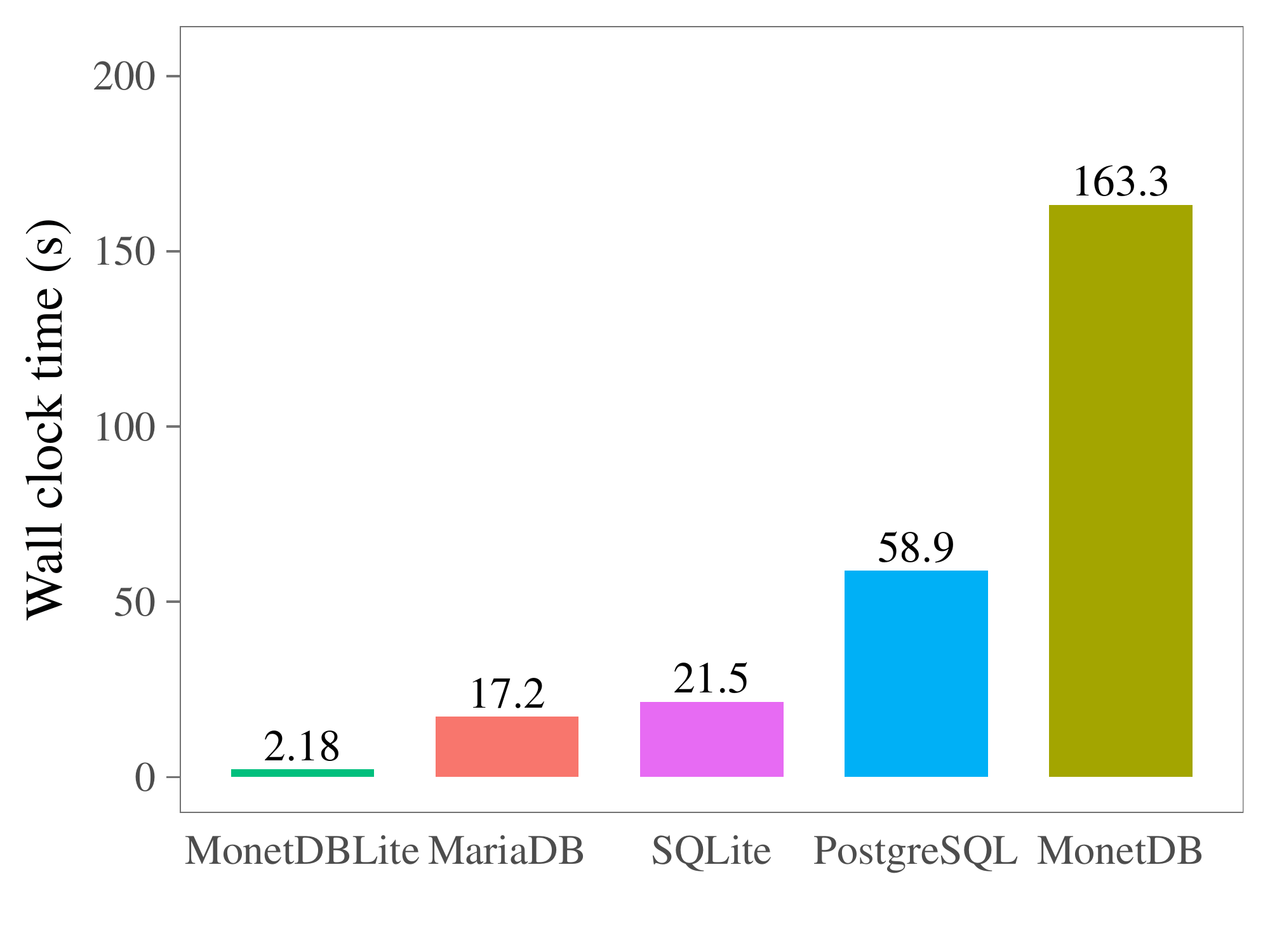}
    \caption{Loading the \texttt{lineitem} into R from the database.}
    \label{fig:dataexport}
\end{figure}

\begin{table*}[!t]
\centering
\begin{tabular}{l|r|r|r|r|r|r|r|r|r|r|r}
\multicolumn{12}{c}{TPC-H SF 1} \\
 System & Q1 & Q2 & Q3 & Q4 & Q5 & Q6 & Q7 & Q8 & Q9 & Q10 & Total \\ 
\hline
MonetDBLite & 0.74 & 0.03 & 0.29 & \textbf{0.06} & \textbf{0.07} & 0.18 & \textbf{0.08} & \textbf{0.08} & \textbf{0.09} & 0.20 & 1.83 \\
MonetDB & 0.87 & \textbf{0.02} & \textbf{0.09} & 0.08 & 0.10 & \textbf{0.05} & 0.08 & 0.11 & 0.16 & \textbf{0.07} & \textbf{1.63} \\
SQLite & 8.41 & 0.04 & 1.83 & 0.44 & 1.00 & 1.17 & 6.52 & {\color{BrickRed}\textbf{T}} & 19.05 & 1.35 & {\color{BrickRed}\textbf{T}}+39.80 \\
PostgreSQL & 8.93 & 0.25 & 0.71 & 2.08 & 0.46 & 1.06 & 0.62 & 0.60 & 2.31 & 1.40 & 18.42 \\
MariaDB & 19.65 & 1.96 & 4.87 & 0.97 & 4.16 & 2.02 & 2.13 & 6.71 & 18.12 & 15.67 & 76.25 \\
data.table & \textbf{0.45} & 0.12 & 0.28 & 0.20 & 0.46 & 0.13 & 0.27 & 0.24 & 0.88 & 0.20 & 3.23 \\
dplyr & 0.70 & 0.13 & 0.34 & 0.25 & 0.60 & 0.17 & 0.31 & 0.41 & 1.17 & 0.28 & 4.37 \\
Pandas & 0.85 & 0.19 & 0.49 & 0.41 & 0.93 & 0.12 & 0.44 & 0.56 & 1.82 & 0.34 & 6.15 \\
Julia & 0.99 & 0.10 & 0.73 & 0.25 & 0.53 & 0.07 & 0.30 & 0.67 & 1.05 & 0.57 & 5.26 \\
\multicolumn{12}{c}{TPC-H SF 10} \\
 System & Q1 & Q2 & Q3 & Q4 & Q5 & Q6 & Q7 & Q8 & Q9 & Q10 & Total \\ 
\hline
MonetDBLite & 16.55 & 0.14 & 1.92 & \textbf{0.50} & \textbf{0.64} & 0.44 & \textbf{0.68} & \textbf{0.75} & \textbf{0.95} & 0.99 & 23.56 \\
MonetDB & \textbf{9.63} & \textbf{0.07} & \textbf{1.15} & 0.87 & 1.16 & \textbf{0.38} & 1.00 & 1.12 & 1.66 & \textbf{0.68} & \textbf{17.69} \\
SQLite & 97.61 & 0.37 & 23.17 & 4.44 & 12.65 & 11.69 & {\color{BrickRed}\textbf{T}} & {\color{BrickRed}\textbf{T}} & {\color{BrickRed}\textbf{T}} & 14.72 & {\color{BrickRed}\textbf{T}}+164.65 \\
PostgreSQL & 88.77 & 2.71 & 63.87 & 22.87 & 4.92 & 11.41 & 7.68 & 6.73 & 74.42 & 63.54 & 346.91 \\
MariaDB & 169.58 & 20.76 & 124.59 & 13.34 & 78.88 & 33.42 & 88.72 & 139.68 & 218.65 & 234.95 & 1122.57 \\
data.table & {\color{BrickRed}\textbf{E}} & {\color{BrickRed}\textbf{E}} & {\color{BrickRed}\textbf{E}} & {\color{BrickRed}\textbf{E}} & {\color{BrickRed}\textbf{E}} & {\color{BrickRed}\textbf{E}} & {\color{BrickRed}\textbf{E}} & {\color{BrickRed}\textbf{E}} & {\color{BrickRed}\textbf{E}} & {\color{BrickRed}\textbf{E}} & {\color{BrickRed}\textbf{E}} \\
dplyr & 31.48 & 1.20 & 5.13 & 3.79 & 8.13 & 1.83 & 4.35 & 4.47 & 16.29 & 3.77 & 80.44 \\
Pandas & {\color{BrickRed}\textbf{E}} & {\color{BrickRed}\textbf{E}} & {\color{BrickRed}\textbf{E}} & {\color{BrickRed}\textbf{E}} & {\color{BrickRed}\textbf{E}} & {\color{BrickRed}\textbf{E}} & {\color{BrickRed}\textbf{E}} & {\color{BrickRed}\textbf{E}} & {\color{BrickRed}\textbf{E}} & {\color{BrickRed}\textbf{E}} & {\color{BrickRed}\textbf{E}} \\
Julia & 24.61 & 5.00 & 7.32 & 2.78 & 9.51 & 0.66 & 7.32 & 13.42 & 18.90 & 6.14 & 95.66 \\
\end{tabular}
\vspace{0.2em}
\caption{Performance Results for TPC\-H SF1 and SF10}
    \label{tbl:tpch}
\end{table*}

For the data export benchmark, we again only consider the \texttt{lineitem} table of the TPC-H benchmark. For this experiment, we read the entire \texttt{lineitem} table from the database into R using the \texttt{dbReadTable} function of the R DBI. This effectively performs a \texttt{SELECT * FROM lineitem} query on the database and stores the result of this query inside an R data frame.

The results of this experiment can be seen in Figure~\ref{fig:dataexport}. We can see that MonetDBLite has by far the fastest data export rate. Because it runs within the analytical process itself, and because it makes use of zero-copy data transfer of numeric columns, the data can be transferred between the database system and R for almost no cost. By contrast, the databases that are connected through a socket connection take a significantly longer time to transfer the result set to the client.

Despite running in-process as well, SQLite also takes a very long time to transfer data from the database to the analytical tool. This is because the conversion of data from a row-major to column-major format takes a significant amount of time. 

\subsection*{Query Execution}
For the query execution benchmark, we run the first ten queries of the TPC-H benchmark inside each of the systems. For each of the libraries, we have created an equivalent script for each of the queries using each of the libraries. 

\textbf{Library Implementations.} Note that, since the libraries naively execute user code without performing any high-level strategic optimizations, there is a lot of room for modifying their performance as the equivalent functionality could be implemented in many naive and inefficient ways. In the worst case, we could perform cross products and filters instead of performing standard joins. Likewise, we could choose poor join orders or not perform filter or projection push down, and force materialization of many unused tuples.

To attempt to maximize the performance of these libraries, we manually perform the high-level optimizations performed by a RDBMS such as projection pushdown, filter pushdown, constant folding and join order optimization. We have created these implementations by using the query plans that are executed by VectorWise~\cite{vectorwise}, a state--of--the--art analytical database system that is the front runner on the official TPC-H benchmark for single node machines. All the scripts that we have created for each of the libraries can be found in our software repository. We have also reached out to the developers of each library and received feedback on optimization.

However, having the user apply all these optimizations is not realistic. This scenario assumes the user has perfect knowledge on how to order joins and assumes the user does not do any inefficient steps such as including unused columns. The benchmark results provided for these libraries should therefore be seen as a \emph{best-case performance scenario}. The benchmark results for these libraries would be significantly worse if we did not manually perform many of the automatic optimizations performed by a database system.

\subsection*{TPC-H SF1}
The total time required to complete all the measured TPC-H queries for the different systems is shown in Table~\ref{tbl:tpch}. We can see that both MonetDB and MonetDBLite show the best performance on the benchmark. They also show very similar performance. This is because the TPC-H benchmark revolves around computing aggregates, and does not involve transferring a large amount of data over the socket connection. As such, the bottleneck is almost entirely the computation performed in the database server. As MonetDB and MonetDBLite use the same internal query execution engine, they have identical performance.

After MonetDB, we can see the various libraries we have tested performing similarly with only a factor two difference between the best and the worst performing library. The fastest library, \texttt{data.table}, is heavily optimized for performing efficient relational operations. However, even with the optimizations we have performed on the user code it still cannot reach the performance of an actual analytical database system. This is because the procedural nature of these libraries heavily limits the actual optimizations that can be performed compared to the optimizations that a database can perform on queries issued in the declarative language of SQL. For example, they do not perform late materialization.

The traditional database systems perform significantly worse than the libraries, however. As the TPC-H benchmark is designed to operate on large chunks of a subset of the columns of a table, the row-store layout and tuple--at--a--time processing methods of the traditional database systems perform extremely poorly on this benchmark. We can see that the traditional database systems perform many orders of magnitude worse than the analytical database systems and the libraries we have used.

\textbf{Individual Query Performance.} The performance of each of the systems on individual queries can be seen in the table as well. The libraries perform extremely well on TPC-H Query 1 and Query 6. On Query 1, \texttt{data.table} even manages to beat our analytical database system. The libraries perform well on these queries because the queries only involve performing filters and aggregations on a single table without any joins.

The libraries perform worse on queries involving multiple joins. The join operations in these libraries do not take advantage of meta-data and indices to speed up the joins between the different tables. As such, they perform significantly worse than the analytical database even when using an optimal join order.

The traditional database systems perform poorly on queries that involve a lot of tuples behind pushed through the pipeline to the final aggregations. Because of their tuple--at--a--time volcano processing model they invoke a lot of overhead for each tuple that passes through the pipeline. This results in poor performance when many tuples have to be processed at a time.

\begin{figure}[!t]
    \includegraphics[width=.8\columnwidth]{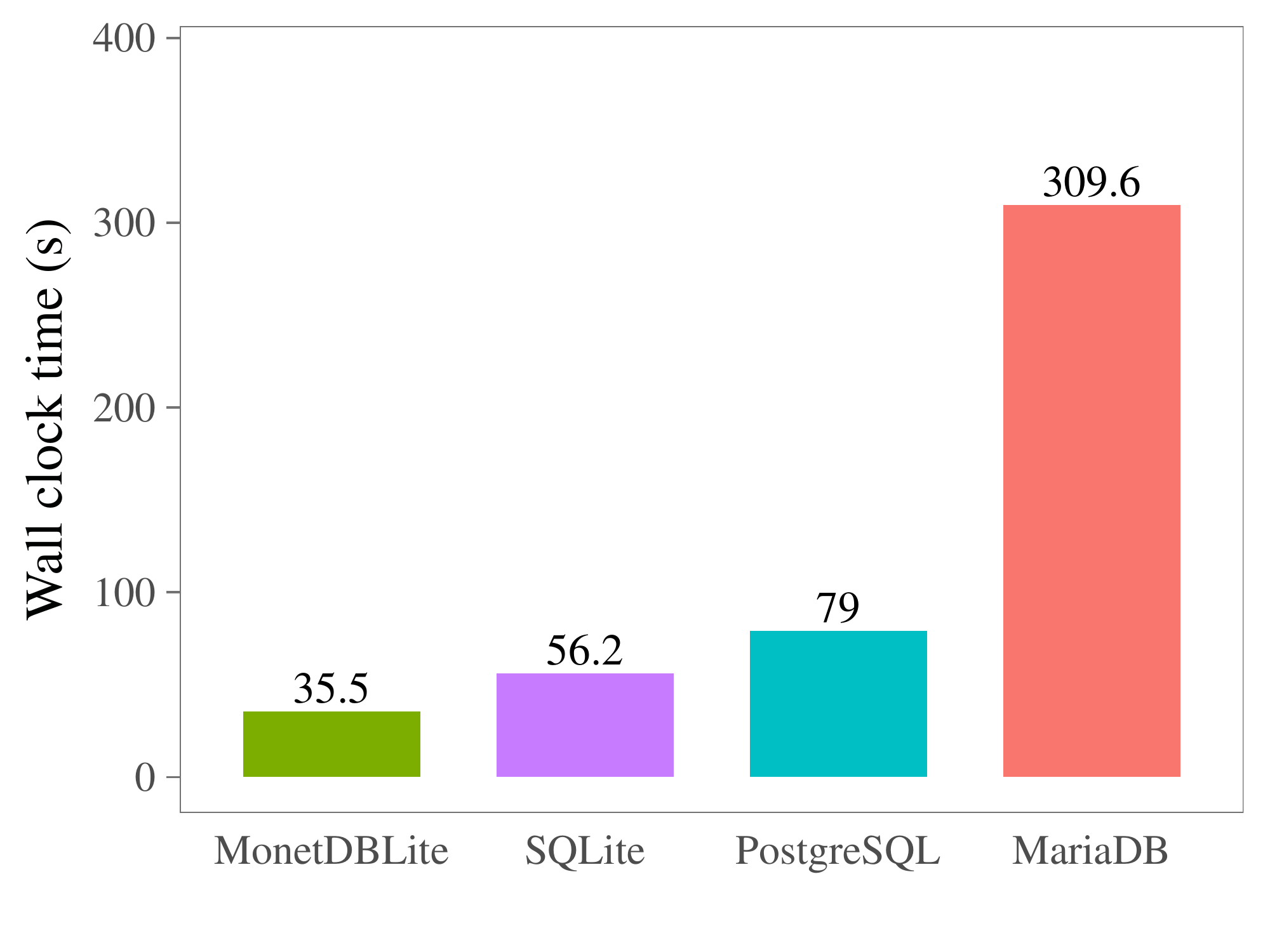}
    \caption{Loading the ACS data into the database.}
    \label{fig:acsload}
\end{figure}

\subsection*{TPC-H SF10}
The results for the TPC-H SF10 benchmark are shown in Table~\ref{tbl:tpch}. We note that at this scale factor, the entire dataset still fits in memory. However, each of the scripting libraries run into either out--of--memory errors or heavily penalized performance from swapping on these queries. This is because these libraries require not only the entire dataset to fit in memory, but also require any intermediates created while processing to fit in memory. When the intermediates exceed the available memory of the machine the program crashes with an out--of--memory exception. The database solutions do not suffer from this problem, as they offload unused data to disk using either the buffer pool or by letting the operating system handle it using memory mapped files.

While the traditional database systems do not run into crashes due to running out--of--memory, their performance does degrade by more than an order of magnitude. Because of the row-store layout of these systems, they have to scan and use the entire dataset rather than only the hot columns. As a result, they run into performance penalties as the entire dataset plus the constructed indices do not fit in memory anymore and have to be swapped to disk. The column-store databases do not suffer from this problem because only the actually used columns have to be touched to answer the queries, and these are small enough to be kept in memory.

\begin{figure}[!t]
    \includegraphics[width=.8\columnwidth]{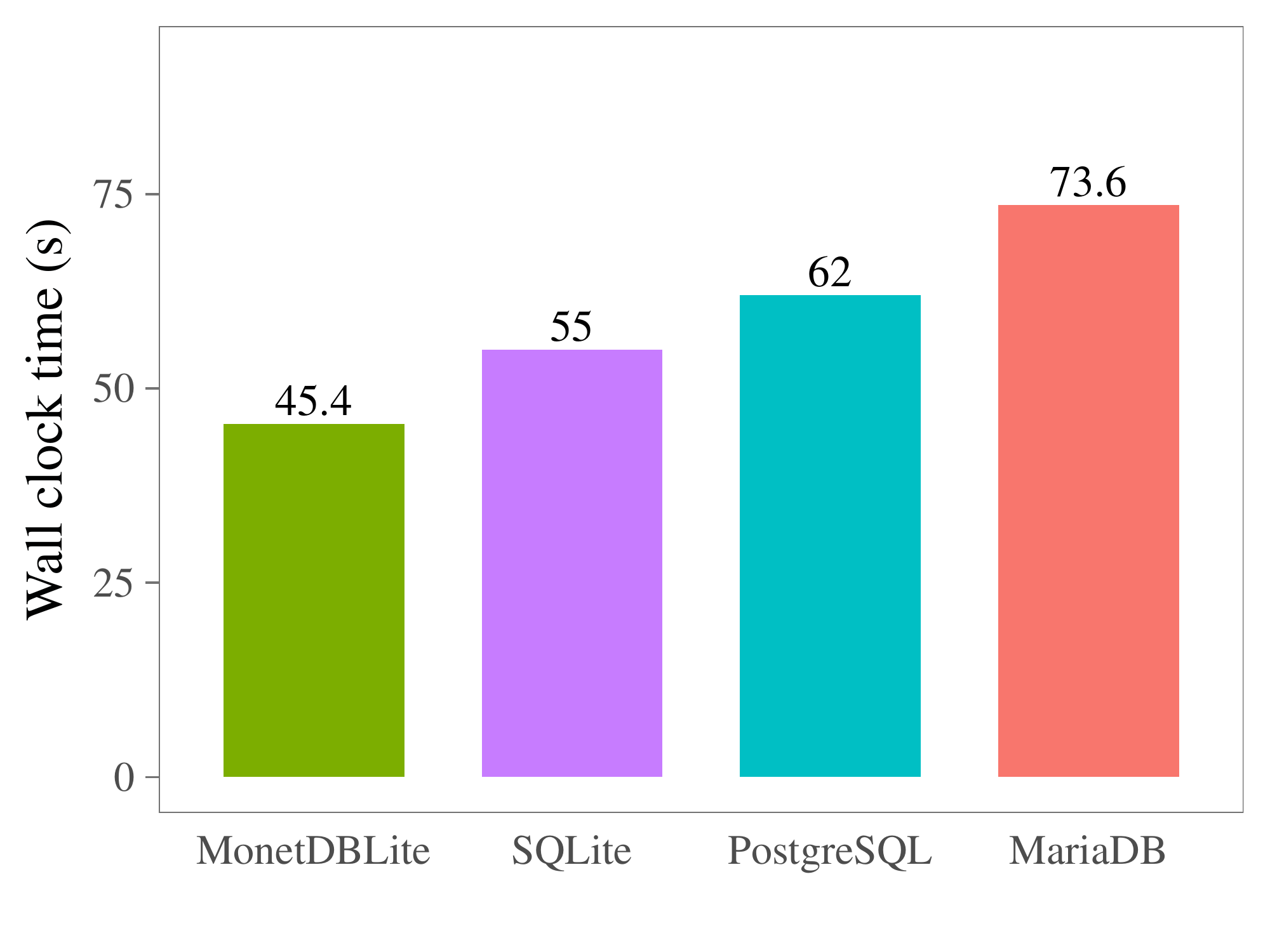}
    \caption{Performing the ACS statistical analysis.}
    \label{fig:acsbenchmark}
\end{figure}

\subsection{ACS Benchmark}
For the American Community Survey benchmark, we run the ACS survey analysis script as provided by Anthony Damico~\cite{acsbenchmark}. The script in this benchmark wrangles data of the American Community Census, a large scale census performed in the United States that gathers data about roughly 1\% of the US population every year.

The script consists of two phases. In the first phase, the required data is gathered and downloaded from the official data repositories. In the second phase, the downloaded data is then processed and stored persistently in a database server. The persistently stored data can then be analyzed and various aggregations and statistics can be gathered from the data using the survey package~\cite{survey}. 

The survey package allows you to hook your own database driver into the script, and will perform a significant amount of processing inside the database. For operations were SQL is insufficient, the data is transferred from the database to R and the data is then processed inside R using various statistical libraries.

The official documentation of the ACS script describes a large amount of statistics that can be gathered from the data. For this benchmark, we benchmark both the required loading time into the database (but exclude the time spent on downloading the data) and a number of statistical operations that are described in the official documentation. We limit ourselves to a subset of the data: we only look at the data from five states of the year 2016. This is $\approx 2.5$ GB in data.

\subsection*{Data Loading}
The benchmark results for loading the data in the database are shown in Figure~\ref{fig:acsload}. MonetDBLite performs the best on this benchmark, but not by as large a factor as seen in the TPC-H benchmark. This is because the survey package performs a lot of preprocessing in R that happen regardless of which database is used. As a result, the performance difference between the different databases is not as overwhelming but still very visible.

\subsection*{Statistics}
The benchmark results for running the various statistical functions using the different database connectors are shown in Figure~\ref{fig:acsbenchmark}. We can see that the difference between performance of the different database engines is not very large. This is because most of the actual processing happens inside R rather than inside the database. The observed difference in performance is mainly because of the difference in the cost of exporting data from the database. However, since the amount of exported data is not very large compared to the amount of processing that occurs in this scenario there is less than a factor two difference between the systems.

\section{Conclusion}\label{section:conclusion}
In this paper, we have presented the embedded analytical database system MonetDBLite. MonetDBLite performs orders of magnitude better than traditional relational database systems when executing analytical workloads, and provides an order of magnitude faster interface between the database and the analytical tool.

In addition to being significantly faster, MonetDBLite is also easier to setup and use because it does not require an external server and does not have any dependencies. It can be installed through standard package and library managers of popular analytical tools. All of these factors combined make MonetDBLite highly suitable as a persistent data store for analytical tasks.

\subsection{Future Directions}
There are still several open issues that result from the nature of how MonetDBLite was created. Because the database that is based on, MonetDB, operates as a stand-alone server several limitations are present in the code that introduce problems when it is used as an embedded database.

MonetDB traditionally only allows a single database process to read the same database. There are is no fine grained locking between several database processes. Instead, a global lock is used on the entire database. If the user attempts to start a database server with a database that is currently occupied by another server an error will be thrown (``database locked'') and the process will exit. This makes sense in the stand-alone server scenario, as running multiple database servers on the same database does not make much sense. However, it is a problem in the embedded database scenario because multiple processes might want to access the same database independently.

Another limitation is that the MonetDB server can only run on a single database at a time because of the large amount of global variables present in the codebase of MonetDB. This is no problem in the stand-alone server case, because another server can be started in a different database directory. However, for the embedded case, this is a painful limitation because only a single database can be opened in the same process.


\balance

\subsection*{Acknowledgments}
We thank the MonetDB contributors at the CWI Database Architectures group for their continued help. We thank Anthony Damico for his help on reporting and resolving bugs and his help on integrating MonetDBLite into his survey analysis scripts. We also thank Pedro Ferreira for his work integrating MonetDBLite into the JVM. This work was funded by the Netherlands Organisation for Scientific Research (NWO), project ``Process mining for multi-objective online control'' (Raasveldt) and ``Capturing the Laws of Data Nature'' (M\"uhleisen).

\bibliographystyle{ACM-Reference-Format}
\bibliography{acmart}
\balance

\end{document}